\documentclass[11pt]{article}
\usepackage{moriond,epsfig}
\usepackage{hyperref}

\def\be{\begin{equation}}
\def\ee{\end{equation}}
\def\bea{\begin{eqnarray}}
\def\eea{\end{eqnarray}}

\begin{document}
\vspace*{4cm}
\title{A VECTOPHOBIC 2HDM IN THE LIGHT OF THE LHC}

\author{ Elvira Cerver\'{o} \footnote{Talk given at \textit{Rencontres de Moriond, EW Interactions and Uniﬁed Theories}, March 3rd-10th, 2012, La Thuile, Italy.} }

\address{Centre for Cosmology, Particle Physics and Phenomenology (CP3) UCLouvain,\\
Chemin du Cyclotron 2, B-1348, Louvain-la-Neuve, Belgium.}

\maketitle\abstracts{We consider recent LHC flavour and Higgs results to constraint a custodial vectophobic Two-Higgs-Doublet-Model (2HDM) where the Minimal Flavour Violation principle (MFV) is implemented to suppress (pseudo)scalar induced flavour changing neutral currents (FCNC).}

\section{Introduction}

For decades the Standard Model (SM) has been the theoretical paradigm in the understanding of fundamental interactions and elementary particles. Yet, its predicted scalar boson still remains one of the key predictions not yet observed experimentally. Current experiments, such as those taking place at the Tevatron and the LHC colliders, are providing the community with interesting data and constraints on the parameter space of many models beyond the SM. In such a perspective, this work \cite{Cervero:2012cx} analyzes a model with two scalar doublets where SM accidental symmetries are minimally violated.

\section{Accidental Symmetries}

Accidental symmetries in the SM are known to provide interesting features that successfully account for numerous experimental results.

The custodial symmetry associated to the $SU(2)_{L+R}$ group arises from the SM scalar potential and appears to be responsible for the success of the tree-level mass relation \cite{Sikivie:1980hm}
\begin{equation}
\rho=\frac{M^2_W}{M^2_Z\cos^2\theta_W}=1.
\end{equation}

\noindent Corrections to the $\rho$ parameter are generated by the sectors where the custodial symmetry is not conserved, and mostly by the Yukawa couplings due to the huge bottom-top mass splitting. These corrections are rather small and compatible with the current experimental measurements.

In general 2HDM scenarios, the custodial symmetry is no longer present in the scalar potential and corrections to the $\rho$ parameter are expected to be too large. However, this symmetry can be imposed. An example is given by a particular 2HDM where the physical states can be classified according to custodial multiplets in the following ways \cite{Gerard:2007kn}
\begin{equation}
\begin{array}{ccc}
\Phi_1 \ni\left\{\begin{array}{ccc}G^{+}\\G^0\\G^-\end{array}\right\}\oplus\{h^0+\frac{v}{\sqrt 2}\}	&	&\Phi_2 \ni\left\{\begin{array}{ccc}H^{+}\\A^0\\H^-\end{array}\right\}\oplus \{H^0\}\mbox{  or  }\left\{\begin{array}{ccc}H^{+}\\H^0\\H^-\end{array}\right\}\oplus \{A^0\}
\end{array}.\label{doublets}
\end{equation}

\noindent In the previous equation $\Phi_1$ is the SM-like doublet with a non-zero vacuum expectation value and a SM-like scalar while $\Phi_2$ contains all the new physical states. In the following we will consider two possible hierarchies corresponding to the case where the singlet custodial component of $\Phi_2$ is the lightest particle, i.e., either the scalar $H^0$ or the pseudoscalar $A^0$ are light.\\

The flavour symmetry \cite{Chivukula:1987py} is another accidental symmetry of the SM. It is associated to the group $G_f=SU(3)_{Q_L}\times SU(3)_{U_R}\times SU(3)_{D_R}$ and is also broken by the Yukawa couplings that give rise to mass terms. In fact, the misalignment between the weak eigenstates and the mass eigenstates induces flavour changing charged currents weighted by the Cabibbo-Kobayashi-Maskawa (CKM) mixing matrix.

In a 2HDM, however, new sources of flavour violation are introduced by the couplings between the fermions and the second doublet
\begin{equation}
\mathcal{L}_{Yukawa}=-\bar Q'_L(Y'_d\Phi_1+Z'_d\Phi_2)d'_R-\bar Q'_L(Y'_u\tilde{\Phi}_1+Z'_u\tilde{\Phi}_2)u'_R,\label{yukawa2HDM}
\end{equation}

\noindent giving rise to FCNC. Following the 2HDM defined in eq. (\ref{doublets}) the FCNC are mediated by the additional neutral states $H^0$ and $A^0$. In order to reduce but not eliminate these FCNC we introduce the MFV hypothesis \cite{D'Ambrosio:2002ex} \cite{Smith:2009hj}. To do so, the SM-like Yukawa couplings $Y_{d,u}$ are promoted into auxiliary fields or spurions and flavour structures can be constructed out of them in a $G_f$ invariant way. Once the spurions are frozen to their background values, the new flavour structures will be expressed in terms of the SM masses and mixing parameters. To apply this formulation of MFV to eq. (\ref{yukawa2HDM}) one just needs to write $Z_{d,u}$ as series in terms of $Y_{d,u}$ in a $G_f$ invariant way
\begin{equation}
	 Z'_d = \{\delta_0  + \delta_1Y'_uY'^{\dag}_u + \delta_2(Y'_uY'^{\dag}_u)^2\}Y'_d,\label{ZdMFV}	\qquad Z'_u = \{\upsilon_0 + \upsilon_1Y'_uY'^{\dag}_u + \upsilon_2(Y'_uY'^{\dag}_u)^2\}Y'_u.
\label{ZuMFV}\end{equation}

\noindent Here, the down Yukawa couplings are neglected in the MFV expansion. Once the $Y_{d,u}$ couplings are diagonalized, the flavour changing couplings, appearing only in the down-sector, will be dependent on non-diagonal CKM elements and will therefore be suppressed. The $\delta_i$ and $\upsilon_i$ coefficients are assumed to be of order one.

\section{Flavour constraints}

\begin{figure}[htbp]
\begin{tabular}{cc}
\includegraphics[scale=0.27]{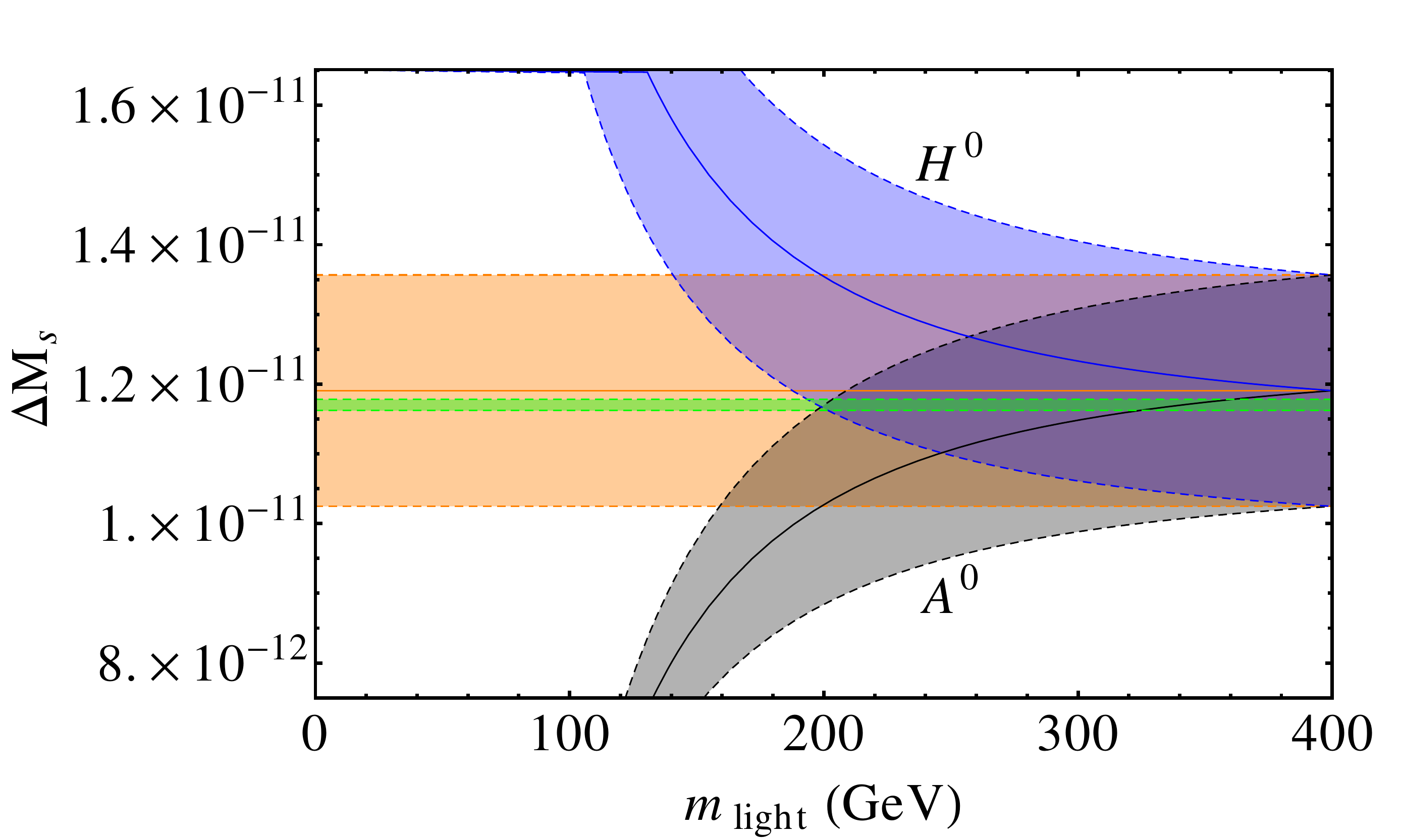}	&\includegraphics[scale=0.26]{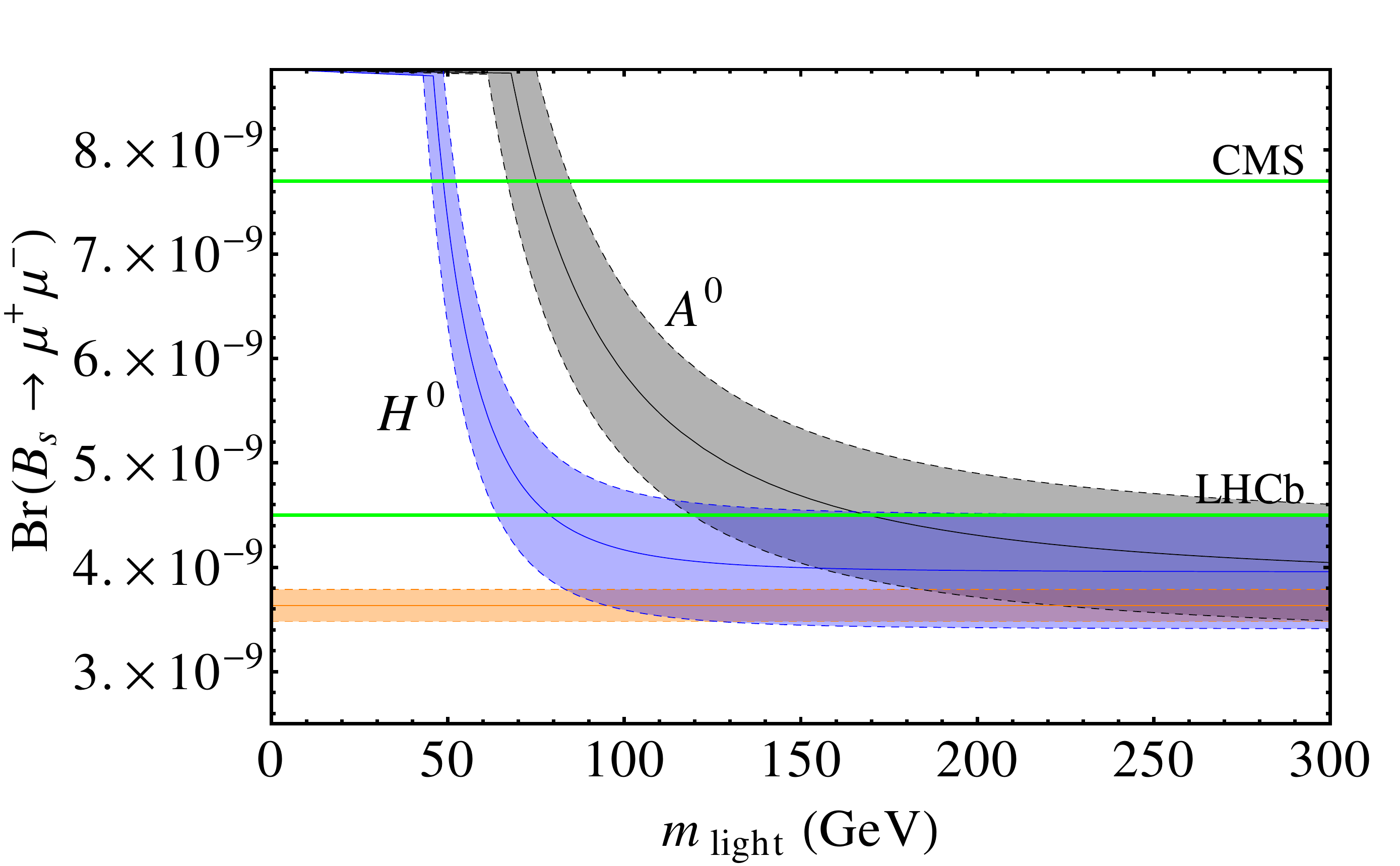}	
\end{tabular}
 \caption{\small{\textit{$\Delta M_{B_s}$ and $B_s\to\mu^+\mu^-$ branching ratio as a function  of the $H^0$ and $A^0$ masses for a MFV coefficient $\delta_1=1$. The green lines indicate the experimental values, the orange areas indicate the 1$\sigma$ SM prediction, the blue areas show the 1$\sigma$ prediction for $m_{H^0}\ll m_{A^0}=400$ GeV, while the grey areas correspond to the analogue prediction for $m_{A^0}\ll m_{H^0}=400$ GeV.}}}\label{neutralmesons}\end{figure}

Consequences of the MFV suppressed FCNC have been studied \cite{Cervero:2012cx} for $\Delta F = 2$ transitions. At the hadronic level, for the $K^0$ and $B^0_q$ systems, the matrix element of the effective Hamiltonian is given by
\begin{equation}
\langle \bar M^0|\mathcal H^{\Delta F =2}_{eff}|M^0\rangle\simeq\langle \bar M^0|\mathcal H^{\Delta F =2}_{eff}|M^0\rangle^{SM}\left[1+16\pi^2 x \delta^2_1 m^2_M\left(\frac{1}{m^2_{H^0}}-\frac{1}{m^2_{A^0}}\right)\right],
\label{matrixelement}\end{equation}

\noindent in the limit $m_{d_j}\ll m_{d_i}$ and with $x$ encoding the full dependence on the top quark mass ($m_t(m_t)=163.4$ GeV)
\begin{equation}
x=\frac{2m_t^4}{M^2_Wv^2S_0(x_t)}\approx 1.61.
\end{equation}
This new contributions to $\Delta F = 2$ transitions could in principle be sizeable. In figure \ref{neutralmesons} (left) the effects of the scalar mediated transitions contributing to the $\Delta F=2$ quantitie $\Delta M_{B_s}$ are shown as a function of the (pseudo)scalar $(A^0)H^0$ mass. In kaon physics, however, the contributions due to neutral Higgses heavier than 100 GeV are totally negligible.

Similarly, the effects of Higgs mediated FCNC on the $B_s\to\mu^+\mu^-$ branching ratio give 
\begin{equation}
\mathcal{B}(B_s\to\mu^+\mu^-)= \mathcal{B}(B_s\to\mu^+\mu^-)^{SM}\left[\left(1+m^2_{B_s}\frac{C_P}{C_A}\right)^2+\left(1-\frac{4m^2_{\mu}}{m^2_{B_s}}\right)m^4_{B_s}\frac{C_S^2}{C^2_A}\right]\label{branching}
\end{equation}

\noindent where $C_A=2Y(x_t)\approx 2.0$ is the Wilson coefficient associated to the SM contribution and the coefficients $C_S$ and $C_P$ are defined by
\begin{equation}
C_{S(P)}=\frac{\Delta}{m^ 2_{H^0(A^0)}};\qquad \Delta = \frac{4\pi^2 \delta_1\lambda_0m^2_t}{M^2_W}.
\end{equation}

The dependence of the $B_s\to\mu^+\mu^-$ branching ratio in terms of the $H^0$ and $A^0$ masses is displayed in figure \ref{neutralmesons} (right).
\section{Diphoton signal at the LHC}
The diphoton signal at the LHC has also been studied considering the possibility that an excess at around 125 GeV \cite{ATLAS:2012ad} \cite{Chatrchyan:2012tw} might be induced by the scalar $H^0$ or the pseudoscalar $A^0$. In the SM, the diphoton decay takes place through top and W loops. In the model studied in this work, both the scalar $H^0$ and the pseudoscalar $A^0$ are vectophobic (i.e., they have no couplings to two gauge bosons) and the decay takes place only through top loops. The number of events in the diphoton invariant mass spectrum is proportional to the production cross-section times the decay branching ratios. The ratio normalized to the SM rate for $m_{H^0(A^0)}=125$ GeV is given in terms of the MFV coefficients and the top Yukawa coupling by 
\begin{small}\begin{equation}
R=\frac{\sigma\times\mathcal{B}(H^0,A^0\to\gamma\gamma)}{\sigma\times\mathcal{B}(h^0\to\gamma\gamma)^{SM}}\rightarrow \left\{\begin{array}{ccc}
R_{H^0/h^0}(m_{H^0}=125 \mbox{ GeV}) = 0.08 \frac{(\upsilon_0+\upsilon_1y^2_t)^4}{(\delta_0+\delta_1y^2_t)^2}\\
\\
R_{A^0/h^0}(m_{A^0}=125 \mbox{ GeV}) = 0.44 \frac{(\upsilon_0+\upsilon_1y^2_t)^4}{(\delta_0+\delta_1y^2_t)^2}
\end{array}\right.\label{ratio}
\end{equation} \end{small}
\begin{figure}[htbp]
\centering\includegraphics[scale=0.27]{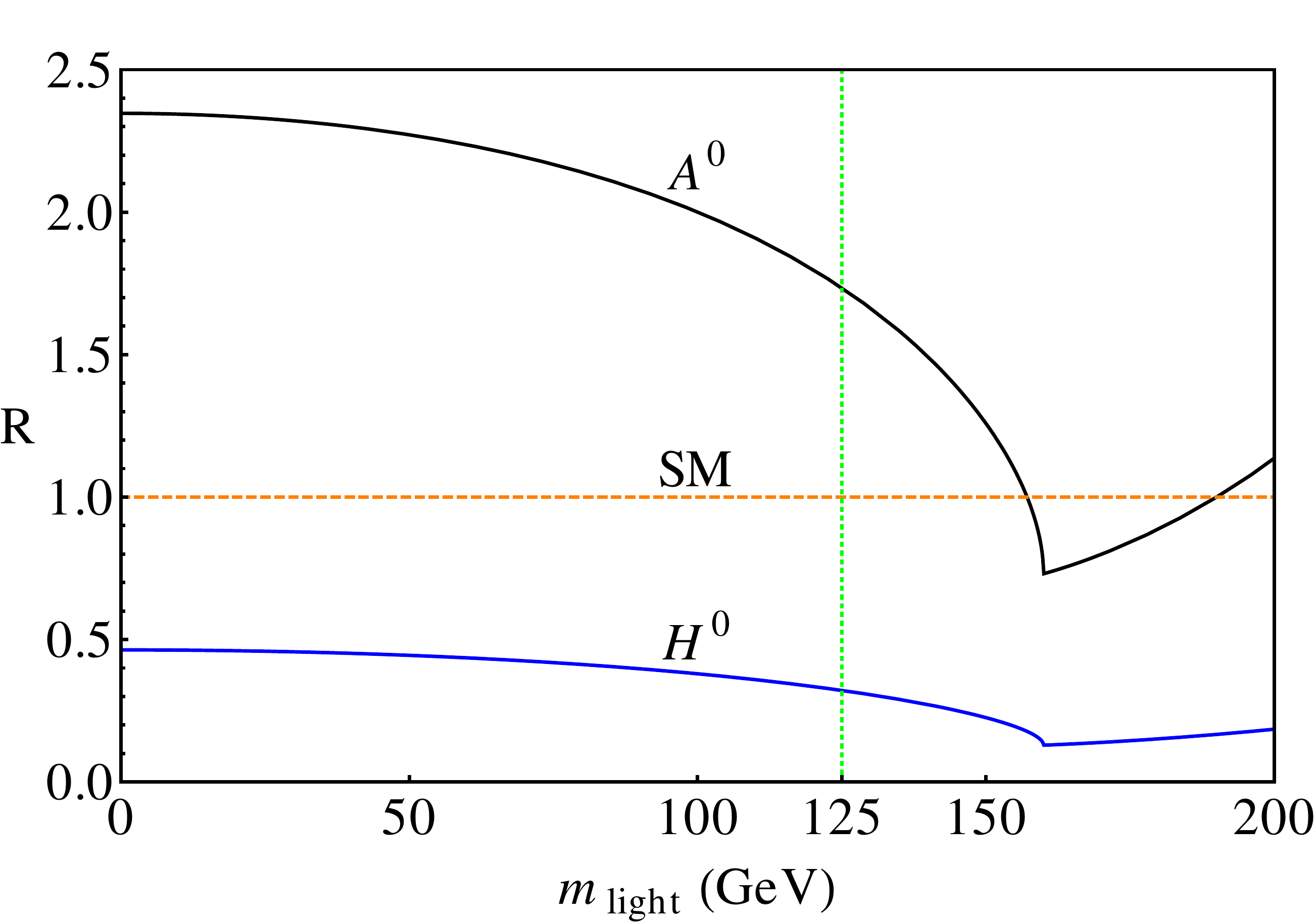}	
 \caption{\small{\textit{The ratio $R$ defined in eq. (\ref{ratio}) as a function of  the $H^0$ and $A^0$ masses if the MFV coefficients are equal to one. The upper (black) curve corresponds to the case where $A^0$ is the lightest non-SM Higgs boson while the lower (blue) one corresponds to  the case where $H^0$ is the lightest non-SM Higgs boson.}}}\label{fig:hto2photons}\end{figure}

In eq. (\ref{ratio}) the production is assumed to be dominated by the gluon-gluon fusion, and the total decay width by the $b\bar b$ final state. The analytical expressions of the ratio R has been used to plot it in terms of the (pseudo)scalar mass as it is shown in figure \ref{fig:hto2photons}. From this plot it can be observed that the pseudo-scalar case might be more suited to account for an observable excess of events in the diphoton channel or even an excess with respect to the SM Higgs boson expectations. It is also clear that both $H^0$ and $A^0$ being vectophobic, any evidence of $W^+W^-$ or $Z^0Z^0$ gauge boson contribution at the production or decay level would require a SM-like scalar. In such a scenario, with two light neutral bosons ($h^0$, and $H^0$ or $A^0$), two resonances in the diphoton invariant mass expectrum would be expected.
\section{Conclusion}
A vectophobic 2HDM with minimal violation of flavour and custodial symmetries has been studied in this work. It has been shown that the $B_s$ system provides us with strong constraints on a light flavour-violating (pseudo)scalar. LHC results in the diphoton invariant mass spectrum have been considered as well in the case of a non SM-like (pseudo)scalar. The light pseudoscalar $A^0$ would be favoured in order to account for a two-photon excess at 125 GeV. However, the additional neutral states of the 2HDM analyzed in this work being vectophobic, more data on the $W^+W^-$ or $Z^0Z^0$ contributions at the production or decay level are needed as well as studies of other channels like the fermionic ones which have attracted special attention after the latest Tevatron results \cite{TEVNPH:2012ab} presented at this workshop.
\section*{Acknowledgments}
I would like to thank the \textit{Rencontres de Moriond EW 2012} organisers for giving me the opportunity to present this work and for their financial support. This work is also supported by the National Fund for Scientific Research (F.R.S.-FNRS) under the FRIA grant. I also thank Jean-Marc G\'{e}rard for the fruitful collaboration and for his relevant comments on this note.

\end{document}